# Social Engagement versus Learning Engagement - An Exploratory Study of FutureLearn Learners

Lei Shi
Department of Computer Science
Durham University
Durham, UK
lei.shi@durham.ac.uk

Alexandra I. Cristea
Department of Computer Science
Durham University
Durham, UK
alexandra.i.cristea@durham.ac.uk

Armando M. Toda
Institute of Mathematics and Computer Sciences
University of São Paulo
São Paulo, Brazil
armando.toda@usp.br

Wilk Oliveira
Institute of Mathematics and Computer Sciences
University of São Paulo
São Paulo, Brazil
wilk.oliveira@usp.br

*Abstract*—Massive Open Online Courses (MOOCs) continue to see increasing enrolment, but only a small percent of enrolees completes the MOOCs. Whilst a lot of research has focused on predicting completion, there is little research analysing the ostensible contradiction between the MOOC's popularity and the apparent disengagement of learners. Specifically, it is important to analyse *engagement not just in learning, but also from a social perspective*. This is especially crucial, as MOOCs offer a growing amount of activities, which can be classified as social interactions. Thus, this study is particularly concerned with how learners interact with peers, along with their study progression in MOOCs. Additionally, unlike most existing studies that are mainly focused on learning outcomes, this study adopts a *fine-grained temporal approach* to exploring how learners progress within a MOOC. The study was conducted on the less explored FutureLearn platform, which employs a social constructivist approach and promotes collaborative learning. The preliminary results suggest potential interesting fine-grained predictive models for learner behaviour, involving weekly monitoring of social, non-social behaviour of active students (further classified as completers and non-completers).

*Keywords—e-Learning, MOOCs, Learning Analytics, Web-based Applications*

## I. Introduction

To date, more than 900 universities deliver over 11,400 courses via Massive Open Online Courses (MOOCs) platforms, attended by about 101,000,000 students [1]. This popularity of MOOCs attracts attention from researchers, practitioners, learners and policy makers worldwide. Understanding this popularity is thus vital for many sectors, if not all, of life. One way is to study what drives students and keeps them engaged, as in this study.

FutureLearn [1], the UK-backed MOOCs platform, has reached the position of 5$^{th}$ in the world, with its 8,700,000 students, hosting around 1,000 courses from universities worldwide [2]. Nevertheless, due to its relative recency (founded in 2012), studies of FutureLearn, as targeted in this study, are much scarcer, compared to other platforms such as edX[2], Udacity[3] and Coursera[4]. FutureLearn employs a social constructivist approach, inspired by Laurillard's Conversation Framework [3], which describes a general theory of effective learning through conversations (or social interactions). Thus, to study learners' engagement on FutureLearn, we explicitly target with this study two types of engagement: *learning engagement* and *social engagement*, by analysing, respectively, *learning activities* and *social activities*, at a *fine-granularity level*. Our research questions are thus:

1. *How do learners' learning and social activities change along the course?*
2. *Are there mutual predictive relationships between learning and social activities?*

## II. Related Work

Exploring engagement in MOOCs has been gaining interest [4]–[7], although there is still no clear collective understanding on the methods of monitoring and measuring it [8], [9]. There have been studies conducted to explore the relationships between learning and social engagement. For example, [10] compared social and learning activity patterns among different demographical groups of students; [11] focused on providing statistical evidence of learner interactions within a MOOC. Results suggested the course instructor's participation in discussions is associated with a higher level of learner interactions, e.g. longer discussion threads. However, it did not stop the decline of *learner engagement* along the course. Our study is focused not only on learners' activities, but additionally explores the potential of *mutual prediction* of *learning activities* and *social activities*, as well as *learners' progression* in the course.

Another interesting study [12] explored learners' motivation to learn in MOOCs and found a positive relationship between their learning and social activities, suggesting that the more comments they wrote, the higher their motivation. Although the study clustered learners based on their activities, it did not estimate the appropriateness to use predictive models, or tied the motivation to the learning gains achieved, as in our study.

Similarly, [13] focused on using learners' motivation to predict learning activities. Their model connected attempts in answering quizzes to learning outcomes. Despite of these promising results, the authors did not include social activities in their analysis, as in our current study.

A few more recent studies took into consideration the additional and important dimension of time, e.g., how student engagement changes along a course; how student engagement in the earlier stage of a course may predict their engagement in the later stage and the completion/dropout of the course. For example, [14] analysed temporal quiz solving patterns in MOOCs, aiming at exploring how the first number of weeks of data predicts activities in the last weeks. [15] proposed a light-weight approach to predicting student dropout from MOOCs using the "early engagement pattern", in particular, the access to the learning content pages and the time spent per access, in the first week of a course. [16] examined demographical changes in student subgroups on a weekly time scale, showing that subgroup membership of students may change significantly in the first half of the course, but stabilise in the second half of the course. The current study presented in this paper also considers the dimension of time. Different from the above-mentioned studies, our study is especially focused on the mutual-prediction potential between social engagement and learning engagement, in a finer granularity, i.e., student engagement in each week of a course, rather than the first versus last, or earlier versus later, stages of a course.

---

[1] https://www.futurelearn.com
[2] https://www.edx.org
[3] https://www.udacity.com
[4] https://www.coursera.org

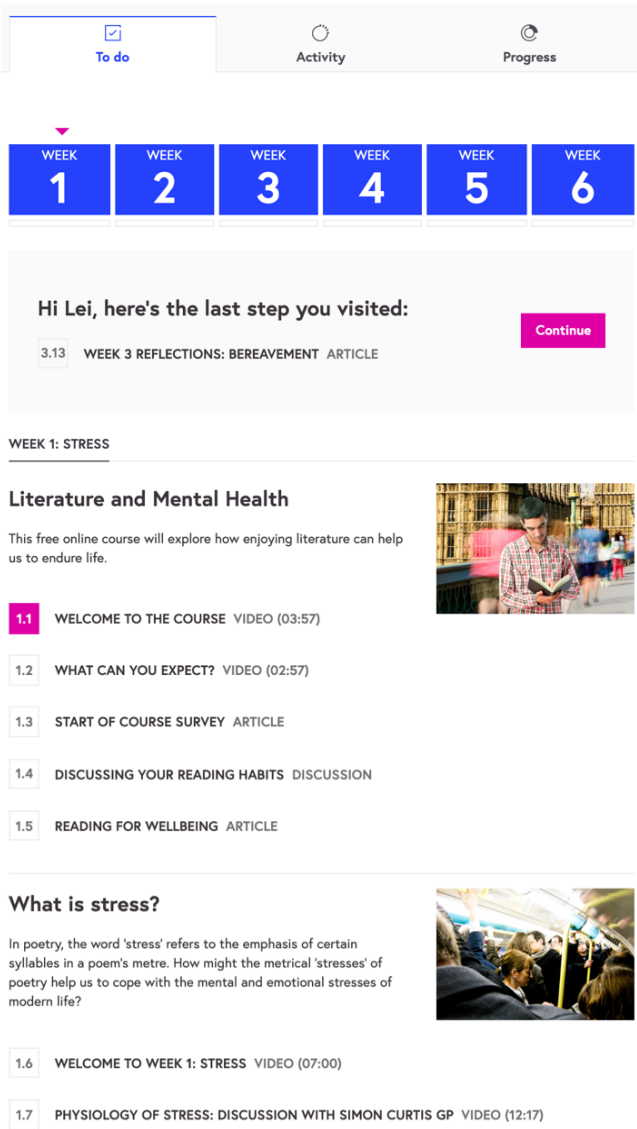

Fig. 1. FutureLearn user interface – the course structure page

## III. FUTURELEARN AND THE COURSE SELECTED

The course explored in this study, "Literature and Mental Health", aims to explain how poems, plays and novels can help people understand and cope with times of deep emotional strain. We specifically chose this course, because, unlike other MOOCs hosted on the platform, where quizzes are also available, this course only supports learning and social activities, which are what this study is focused on.

The course includes six themes, including (1) stress, (2) heartbreak, (3) bereavement, (4) trauma, (5) depression and bipolar, and (6) aging and dementia. It is structured into 6 weekly learning units. Week 1 through Week 6 contains 19, 15, 13, 13, 16 and 18 *steps*, respectively. In total, there are 94 *steps* in the course. *Steps* are basic learning items, which may contain articles, images and videos. Fig. 1 shows the navigation page of the course, where learners can click a blue 'WEEK' button on the top to navigate to a weekly learning unit, or click the *step* title to navigate to the *step* page.

At the start of each week, the learners receive a brief email to introduce the week's topic. They can learn at their own pace, but they are encouraged to join the discussions happening in the current week. When a learner has finished reading an article, or watching a video on a step page, they may click the big pink 'Mark as Complete' button and then click the '>' button, to move on to the next step, as shown in Fig. 2. Marking steps as complete updates their progress page, which helps them keep track of the steps that they have done on the 'To Do' list and makes their progress eligible for a 'Statement of Participation' (learners must mark over 50% of course steps as complete to be eligible, according to the FutureLearn policy [17]). The learners are encouraged to discuss their interests, knowledge, and experiences with their peers throughout the course. They can leave a comment on a step page; they can also reply to or 'like' a comment. To see the discussions, as shown in Fig. 3, they need to click the pink '+' button on the bottom left of the step page (Fig. 2).

The pedagogy of the course is principally dictated by FutureLearn: learners study materials prepared by the course team; materials are chunked into small parts, i.e. steps, within each week; learners are expected to complete the dictated activities, e.g. accessing materials (visiting step pages which may contain articles, images, and videos) and participating in discussions (posting, replying or liking comments) to consolidate their learning and apply their knowledge. These activities provide learners a safe and controlled environment to practice their new skills [18] and obtain experience of new techniques. Quiz or assessment is not available in this course, but there is a reflection step at the end of each weekly learning unit.

Our study refers to three runs of this course. Learners' clickstream data was recorded. The data used in this study was drawn from three sources: (1) enrolments – the enrolment of each learner, including the unique ID assigned to the learner, and whether/when the learner unenrolled from the course; (2) step activity – learning activities which identify when a learner visited and/or completed a step; and (3) comments – social activities, which identify when a learner left a comment. This study was conducted in accordance with the FutureLearn Code of Practice for Research Ethics [5]. All data was completely anonymous, i.e. individual learners were not identifiable.

---
[5] Research Ethics, https://about.futurelearn.com/terms/research-ethics-for-futurelearn

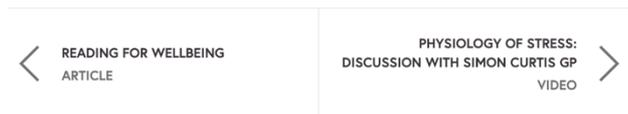

Fig. 2. FutureLearn user interface – a step page

Fig. 3. FutureLearn user interface – the discussion board on a step page

## IV. THE DATASET

The dataset analysed thus contained three runs of the course. In Run 1, there were initially 22,980 learners enrolled, but 2,728 of them proactively unenrolled from the course, with 20,252 learners remaining (88.13%). In Run 2, from 12,285 enrolled, after 1,658 unenrolling, 10,627 (86.50%) remained; in Run 3, the initial 9,479 went down to 8,488 (89.55%). Therefore, in total, out of 22,980 + 12,285 + 9,479 = 44,744 learners enrolled in the course, there were 20,252 + 10,627 + 8,488 = 39,367 (87.98%) remaining learners. Fig. 4 depicts this. We can observe that the course became less popular (with fewer learners enrolled and remaining in consecutive runs).

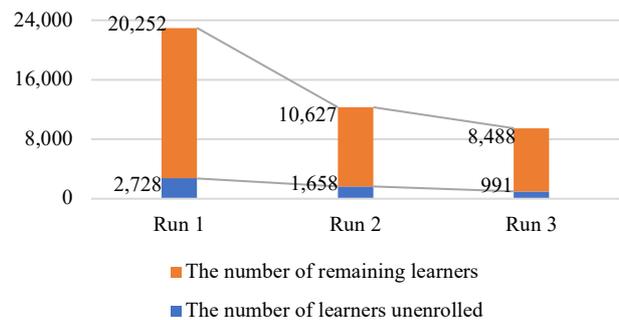

Fig. 4. The number of remaining and unenrolled students in three runs of the course

This study explores *learning engagement* and *social engagement* by analysing *learning activities* and *social activities* of learners, respectively. Table I summarises the activity data analysed in the study. We can observe that the number for each type of activity decreases in consecutive runs. Interestingly, for each run, the number of step completions was very close to the number of step visits, which may be due to the fact that, for this course, at its start, the instructor explicitly emphasised the use of the button "Mark as complete", which is also designed to be conspicuous (Fig. 2).

TABLE I. VISITS, COMPLETIONS AND COMMENTS, IN EACH RUN

|  | **Visits** | **Completions** | **Comments** |
|---|---|---|---|
| *Run 1* | 370,422 | 338,404 | 80,666 |
| *Run 2* | 160,718 | 147,678 | 32,617 |
| *Run 3* | 117,907 | 107,295 | 21,164 |
| **Total** | 649,047 | 593,377 | 134,447 |

## V. METHODOLOGY

To explore learning and social activities along learners' progression in MOOCs, we aggregate the data from all three *runs* and group them into six *weeks*. Additionally, we label *step* visits and completions as *learning activities*, and comments as *social activities*. Table II summarises the numbers of visits (*learning*), completions (*learning*) and comments (*social*), in each week.

TABLE II. VISITS, COMPLETIONS AND COMMENTS, IN EACH WEEK

|  | Week 1 | Week 2 | Week 3 | Week 4 | Week 5 |
|---|---|---|---|---|---|
| *Visits: learning* | 236,571 | 117,436 | 85,212 | 67,607 | 76,618 |
| *Completions: learning* | 210,165 | 109,402 | 79,900 | 63,437 | 70,727 |
| *Comments: social* | 33,989 | 22,614 | 13,128 | 9,243 | 9,679 |

Each week of the course contains a different number of *steps*. The largest number of *steps* in a week is **19** (in Week 1). Thus, to compare activities across weeks in a fair way, we map them onto the existing steps, as follows:

$$\text{visits\_c}(n) = \text{visits}_n \times 19 \div \text{steps}_n \quad (1)$$

$$\text{completions\_c}(n) = \text{completions}_n \times 19 \div \text{steps}_n \quad (2)$$

$$\text{comments\_c}(n) = \text{comments}_n \times 19 \div \text{steps}_n \quad (3)$$

where *n* represents the $n^{th}$ week. For example, in Week 2 (n = 2), there were 117,436 step visits ($\text{visits}_2$ = 117,436), and 15 *steps* ($\text{steps}_2$ = 15), so, the converted number visits_c(2) is 117,436 × 19 ÷ 15 = 148,752. Table 3 shows the converted numbers of learning and social activities, for each week.

TABLE III. CONVERTED NUMBER OF VISITS, COMPLETIONS, COMMENTS, AND LIKES, IN EACH WEEK

|  | visits_c(n) | completions_c(n) | comments_c(n) |
|---|---|---|---|
| *Week 1* (n = 1) | 236,571 | 210,165 | 33,989 |
| *Week 2* (n = 2) | 148,752 | 188,420 | 28,644 |
| Week 3 (n = 3) | 124,541 | 182,021 | 19,187 |
| *Week 4* (n= 4) | 98,810 | 144,415 | 13,509 |
| *Week 5* (n= 5) | 90,984 | 108,043 | 11,494 |
| *Week 6* (n= 6) | 69,175 | 73,018 | 9,620 |

We use these for comparative plots, as can be seen in the next section, and for answering the first research question.

For answering the second question on prediction, we additionally perform normality check (via Kolmogorov–Smirnov test; *p* < .01), and use T-tests for normally distributed data, and Mann-Whitney U tests for the data that is not normally distributed, to compare differences between social activities and learning, across four different categories, as the results show.

## VI. ACTIVITY CHANGES ALONG THE COURSE

In this section, we answer the first research question: ***how do learners' learning and social activities change along the course?***

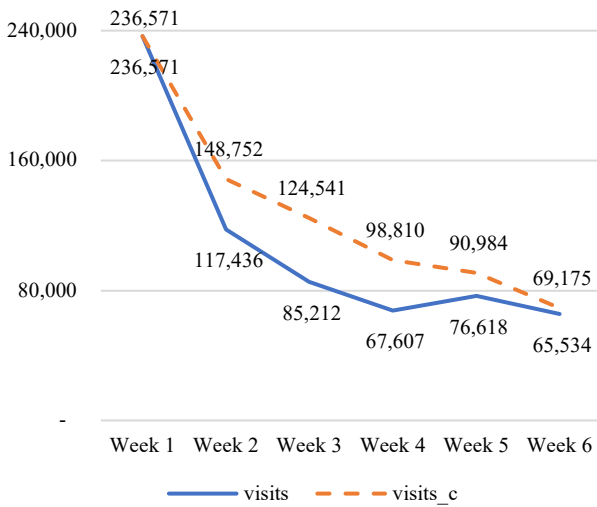

Fig. 5. Number of activities – step visits, across weeks

To compare the activities across weeks, we visualise the data from Table II and Table III into Fig. 5, Fig. 6 and Fig. 7. We can observe that the numbers of learners' activities dropped along the course. There are also another two interesting shared patterns: (1) the converted numbers drop slower than the original numbers; and (2) the converted numbers are raised slightly in Week 5, then drop again in Week 6.

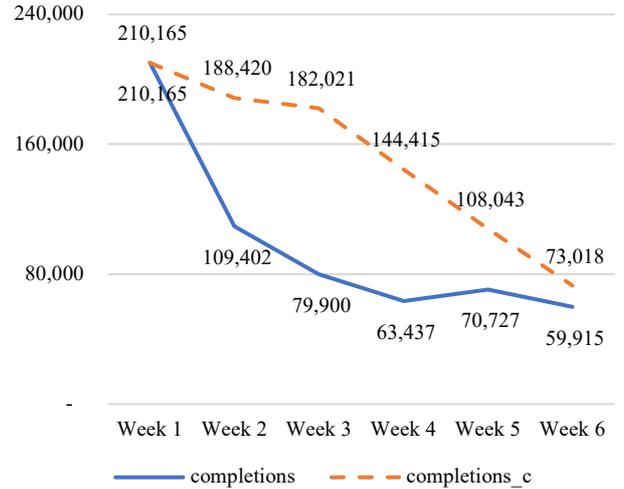

Fig. 6. Number of activities – step completions, across weeks

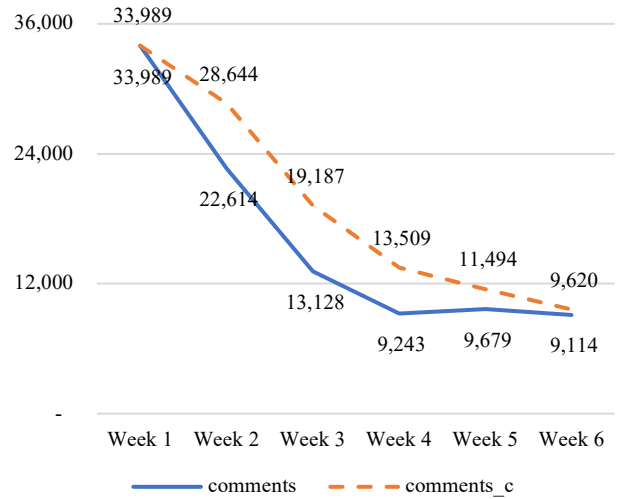

Fig. 7. Number of activities – comments, across weeks

### A. Learner Statistics

In this section, we investigate how many learners performed learning and social activities, in each week, and compare learner numbers across weeks. Fig. 8 shows learning activities, including step visits and completions. Fig. 9 shows social activities, i.e., comments. Overall, for all these three types of activities, the number of learners dropped along the course. The numbers dropped the most from Week 1 to Week 2, then the drop became smoother in the consecutive weeks. These results suggest that the dropouts tend to happen at the very beginning of a course, and once learners have gone through the beginning stage, they are more likely to continue the course.

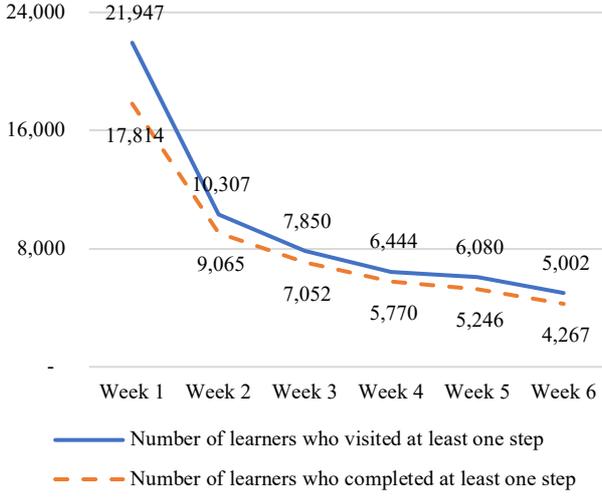

Fig. 8. Number of learners who performed learning activities, across weeks

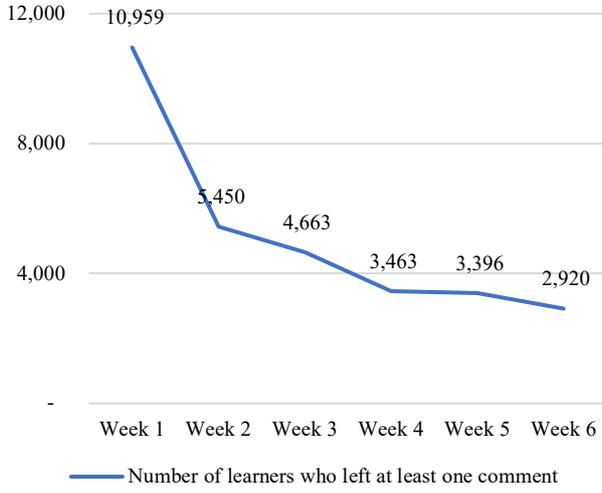

Fig. 9. The number of learners who performed social activities, across weeks

## B. Learner Engagement

Although the overall numbers of activities decrease along the course, it does not necessarily mean that learners became less engaged, because the number of active learners also decreased. Thus, to further investigate how learners were engaged in the course, we divide the numbers of activities (shown in Table III) by the numbers of active learners (shown in Fig. 8 and Fig. 9), using the following three equations:

$$visits\_rate(n) = visits\_c(n) \div learners\_active\_v_n \quad (4)$$

$$completions\_rate(n) = completions\_c(n) \div learners\_active\_comp_n \quad (5)$$

$$comments\_rate(n) = comments\_c(n) \div learners\_active\_comm_n \quad (6)$$

where, in week $n$, $learners\_active\_v_n$ represents the number of learners who visited at least one *step*; $learners\_active\_comp_n$ is the number of learners who completed at least one *step*; $learners\_active\_comm_n$, the number of learners who left at least one comment. Table IV, Fig. 10 and Fig. 11 summarise the statistical results.

TABLE IV. CONVERTED NUMBER OF VISITS, COMPLETIONS, COMMENTS, AND LIKES, IN EACH WEEK

|  | visits_rate(n) | completions_rate(n) | comments_rate(n) |
|---|---|---|---|
| *Week 1* (n = 1) | 10.78 | 11.8 | 3.1 |
| *Week 2* (n = 2) | 14.43 | 20.79 | 5.26 |
| *Week 3* (n = 3) | 15.87 | 25.81 | 4.11 |
| *Week 4* (n = 4) | 15.33 | 25.03 | 3.9 |
| *Week 5* (n = 5) | 14.96 | 20.6 | 3.38 |
| *Week 6* (n = 6) | 13.83 | 17.11 | 3.29 |

From Fig. 10 and Fig. 11, we can observe that, unlike for the actual numbers of activities, as shown in Fig. 5, Fig. 6 and Fig. 7, the peaks of the "rates" appear in the middle of the course, and the troughs are in the first and the last weeks. Interestingly, on the contrary, the lowest points of the "rates" appear in the first week, even if the highest points of actual number of activities appear also in the first week. This suggests that, although the number of activities drop along the course, with the highest drop at the beginning of the course, i.e., from Week 1 to Week 2, the learners who did not drop out from the course tended to be more engaged in both learning and social activities. Therefore, early intervention is crucial for learners' retention in MOOCs.

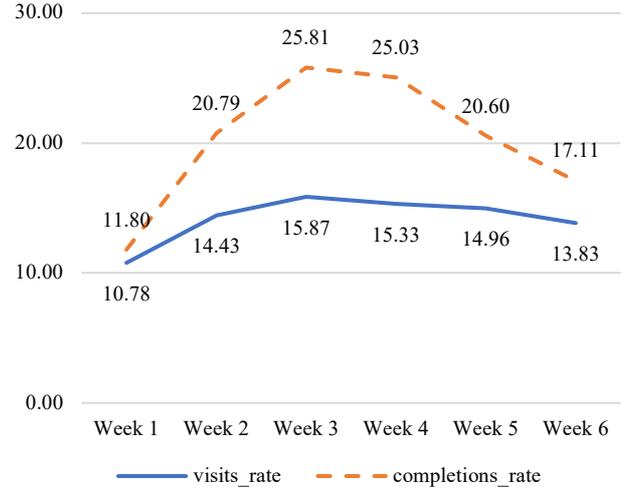

Fig. 10. Learning activities – visit rate and completion rate, across weeks

Comparing "completion rate" with "visit rate", as shown in Fig. 10, the former is much higher than the latter, especially during the middle weeks, i.e., Week 3 and Week 4. This indicates that the learners who felt they completed steps and wanted to mark it in clear, were, overall, more engaged, in terms of performing learning tasks and responding to the platform. Still, even they might become less engaged at the end of the course. Interestingly, the gap between "completion rate" and "visit rate" is the largest during the middle weeks. For example, in Week 3, the gap between these two "rates" is |25.81 – 15.87| = 9.94, being the largest, compared to the gap in Week 1 of 1.02, and in Week 6, 3.28. This indicates that learners who managed to reach the middle stage of a MOOC are most active. However, even they may struggle towards the end. Therefore, during the final stage of MOOCs, it is important to provide learners with additional support, in order to promote course completion.

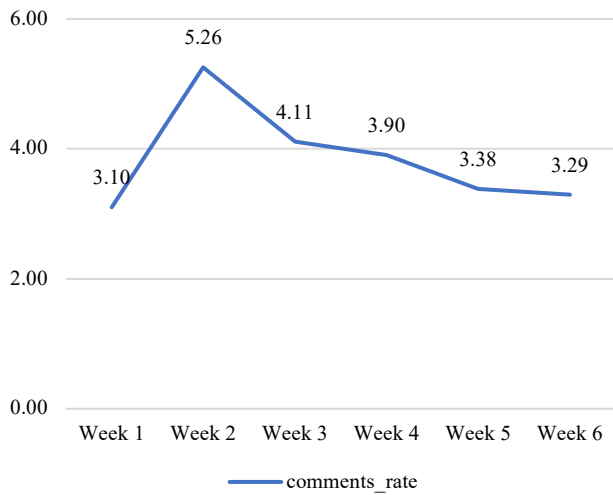

Fig. 11. Social activities - comment rate, across weeks

Overall, comparing Fig. 10 and Fig. 11, we can observe that "rates" of learning activities are much higher than those of social activities, which suggests that, although FutureLearn is strongly built upon the social learning pedagogy [19], social features are less popular than learning features. This may be due to FutureLearn being, first and foremost, a learning platform, rather than a social networking platform. Learners may also be concerned about exposing themselves to their course instructor or peers. Some learners only performed learning activities, e.g., reading or watching learning materials, rather than interacting with their peers, although they might have still been benefitting from reading the comments of their peers. Therefore, when designing MOOCs, it is necessary to keep the balance between learning and social activities: allow a moderate amount of social interaction to increase learners' engagement, but not to overwhelm them.

## VII. PREDICTIVE RELATIONSHIPS BETWEEN LEARNING & SOCIAL ACTIVITIES

In this section, we answer the second research question: *are there mutual predictive relationships between learning and social activities?* To simplify the analysis, we consider only active learners, because most non-active learners did not perform any learning or social activity.

First, we label active learners as Completers and Non-Completers. Completers represent those active learners who completed at least 50% of all steps in the current week (by clicking the button "Mark as complete", as shown in Fig. 2); Non-Completers represent the rest of the active learners. We chose 50% as reference, as it conforms to FutureLearn's definition of "fully participating learners" [17] (since this course does not provide quizzes, quiz completion conditions are not included here). Mann-Whitney U tests were conducted (for the non-normally distributed dataset), for each week, to compare social activities, i.e., comments, between Completers and Non-Completers. The results, as shown in Table V, indicate that, for each week, Completers left significantly (p < .01) more comments than Non-Completers. This suggests that learners who complete more steps may interact more with peers in the **current week**.

TABLE V. COMMENTS (*SOCIAL*): COMPLETERS VERSUS NON-COMPLETERS, IN **CURRENT WEEK**

| | Learner Group | N | Mean Rank | M-W U | Sig. |
|---|---|---|---|---|---|
| *Week 1* | Completer | 12,830 | 31,071.70 | 93,630,810.5 | <.001 |
| | Non-Completer | 31,954 | 18,907.67 | | |
| *Week 2* | Completer | 4,491 | 15,898.39 | 21,833,497.5 | <.001 |
| | Non-Completer | 18,514 | 10,436.80 | | |
| *Week 3* | Completer | 3,757 | 16,367.34 | 17,882,045 | <.001 |
| | Non-Completer | 19,248 | 10,553.53 | | |
| *Week 4* | Completer | 3,087 | 16,062.19 | 16,669,218.5 | <.001 |
| | Non-Completer | 19,918 | 10,796.39 | | |
| *Week 5* | Completer | 2,654 | 16,777.49 | 1,3007,270.5 | <.001 |
| | Non-Completer | 20,351 | 10,815.15 | | |
| *Week 6* | Completer | 2,099 | 17,267.63 | 9,840,898 | <.001 |
| | Non-Completer | 20,906 | 10,924.22 | | |

TABLE VI. STEP VISITS & COMPLETIONS (*LEARNING*): SOCIAL VERSUS NON-SOCIAL LEARNERS, IN **CURRENT WEEK**

| | | Learner Group | N | Mean Rank | M-W U | Sig. |
|---|---|---|---|---|---|---|
| *Week 1* | visits | Social | 4,918 | 38,691.57 | 17,871,647.0 | <.001 |
| | | Non-Social | 39,866 | 20,381.79 | | |
| | completions | Social | 4,918 | 38.835.82 | 17,162,251.0 | <.001 |
| | | Non-Social | 39,866 | 20,364.00 | | |
| *Week 2* | visits | Social | 2,680 | 42,010.46 | 3,843,238.5 | <.001 |
| | | Non-Social | 42,104 | 21,143.78 | | |
| | completions | Social | 2,680 | 40,820.98 | 7,031,038.5 | <.001 |
| | | Non-Social | 42,104 | 21,219.49 | | |
| *Week 3* | visits | Social | 2,380 | 41,605.45 | 4,733,935.0 | <.001 |
| | | Non-Social | 42,404 | 21,314.14 | | |
| | completions | Social | 2,380 | 41,499.74 | 4,985,519.0 | <.001 |
| | | Non-Social | 42,404 | 21,320.07 | | |
| *Week 4* | visits | Social | 1,767 | 42,106.94 | 3,170,104.0 | <.001 |
| | | Non-Social | 43,017 | 21,582.69 | | |
| | completions | Social | 1,767 | 41,928.02 | 3,486,254.0 | <.001 |
| | | Non-Social | 43,017 | 21,590.04 | | |
| *Week 5* | visits | Social | 1,787 | 42,435.54 | 2,600,913.0 | <.001 |
| | | Non-Social | 42,997 | 21,559.49 | | |
| | completions | Social | 1,787 | 42,338.89 | 2,773,623.0 | <.001 |
| | | Non-Social | 42,997 | 21,563.51 | | |
| *Week 6* | visits | Social | 1499 | 42917.48 | 1,675,161.5 | <.001 |
| | | Non-Social | 43285 | 21681.70 | | |
| | completions | Social | 1499 | 42578.24 | 2,183,677.0 | <.001 |
| | | Non-Social | 20,906 | 10,924.22 | | |

Second, we label active learners as Social learners and Non-Social learners. The former label represents those active learners who left at least one comment *and* received at least one like; the latter represents the rest of the active learners. The same method as above, via a Mann-Whitney U tests (see Table VI), shows that, in each week, Social learners visited significantly ($p < .01$) more *steps* and completed significantly ($p < .01$) more *steps*, than Non-Social learners, in each week. This suggests that learners who interact more with their peers may complete more steps in the **current week**.

Third, a Mann-Whitney U test was conducted, to compare Completers and Non-Completers, in terms of comments they left in the following week(s). For example, we compare Completers and Non-Completers in *Week 1*, in terms of comments they left in *Week 2*. The results (see Table VII) show that, for every week, Completers left significantly ($p < .01$) more comments in the following week as well, when compared to Non-Completers. This suggests that learners who complete more *steps* may interact more with their peers in **following week(s)**.

TABLE VII. COMMENTS (*SOCIAL*): COMPLETERS VERSUS NON-COMPLETERS, IN THE **FOLLOWING WEEK(S)**

| | Learner Group | Comments left in the Following Week | | | |
|---|---|---|---|---|---|
| | | N | Mean Rank | M-W U | Sig. |
| Week 1 | Completers | 12,830 | 27,265.91 | 142,459,023.0 | <.001 |
| | Non-completers | 31,954 | 20,435.75 | | |
| | Completers | 4,491 | 15,511.25 | 23,572,149.0 | <.001 |
| | Non-completers | 18,514 | 10,530.71 | | |
| Week 2 | Completers | 3,757 | 15,193.89 | 22,290,679.5 | <.001 |
| | Non-completers | 19,248 | 10,782.58 | | |
| | Completers | 3,087 | 15,973.36 | 16,943,434.5 | <.001 |
| | Non-completers | 19,918 | 10,810.16 | | |
| Week 3 | Completers | 2,654 | 16,111.21 | 14,775,575.5 | <.001 |
| | Non-completers | 20,351 | 10,902.04 | | |
| | Completers | 12,830 | 27,265.91 | 142,459,023.0 | <.001 |
| | Non-completers | 31,954 | 20,435.75 | | |
| Week 4 | Completers | 4,491 | 15,511.25 | 23,572,149.0 | <.001 |
| | Non-completers | 18,514 | 10,530.71 | | |
| | Completers | 3,757 | 15,193.89 | 22,290,679.5 | <.001 |
| | Non-completers | 19,248 | 10,782.58 | | |
| Week 5 | Completers | 3,087 | 15,973.36 | 16,943,434.5 | <.001 |
| | Non-completers | 19,918 | 10,810.16 | | |
| | Completers | 2,654 | 16,111.21 | 14,775,575.5 | <.001 |
| | Non-completers | 20,351 | 10,902.04 | | |
| Week 6* | Completers | n/a | n/a | n/a | n/a |
| | Non-completers | n/a | n/a | n/a | n/a |
| | Completers | n/a | n/a | n/a | n/a |
| | Non-completers | n/a | n/a | n/a | n/a |

* Regarding Week 6, as there were only 6 weeks in the course thus Week 6 did not have "Following Week", the cell in the Week 6's row is n/a.

Finally, Mann-Whitney U tests were also conducted to compare Social with Non-Social learners, in terms of their *step* visit and completion rate in the following week(s). For example, we compare both types of learners in *Week 1*, in terms of their *step* visit and completion rates in *Week 2*. The results (see Tables VIII and IX) show that Social learners visited and completed significantly ($p < .01$) more *steps* in the following week(s) than Non-Social. This suggests that learners who interact more with their peers may visit and complete more *steps* in the **following week(s)**.

TABLE VIII. STEP VISITS (*LEARNING*): SOCIAL VERSUS NON-SOCIAL LEARNERS, IN **FOLLOWING WEEK(S)**

| | Learner Group | Step visits in the Following Week | | | |
|---|---|---|---|---|---|
| | | N | Mean Rank | M-W U | Sig. |
| Week 1 | Social | 4,918 | 34,605.42 | 379,67,372.0 | <.001 |
| | Non-social | 39,866 | 20,885.87 | | |
| | Social | 2,680 | 39,670.38 | 10,114,641.0 | <.001 |
| | Non-social | 42,104 | 21,292.73 | | |
| Week 2 | Social | 2,380 | 40,271.24 | 7,909,359.5 | <.001 |
| | Non-social | 42,404 | 21,389.02 | | |
| | Social | 1,767 | 41,425.47 | 4,374,262.5 | <.001 |
| | Non-social | 43,017 | 21,610.69 | | |
| Week 3 | Social | 1,787 | 41,336.72 | 4,564,491.5 | <.001 |
| | Non-social | 42,997 | 21,605.16 | | |
| | Social | 4,918 | 34,605.42 | 379,67,372.0 | <.001 |
| | Non-social | 39,866 | 20,885.87 | | |
| Week 4 | Social | 2,680 | 39,670.38 | 10,114,641.0 | <.001 |
| | Non-social | 42,104 | 21,292.73 | | |
| | Social | 2,380 | 40,271.24 | 7,909,359.5 | <.001 |
| | Non-social | 42,404 | 21,389.02 | | |
| Week 5 | Social | 1,767 | 41,425.47 | 4,374,262.5 | <.001 |
| | Non-social | 43,017 | 21,610.69 | | |
| | Social | 1,787 | 41,336.72 | 4,564,491.5 | <.001 |
| | Non-social | 42,997 | 21,605.16 | | |

TABLE IX. STEP COMPLETIONS (*LEARNING*): SOCIAL VERSUS NON-SOCIAL LEARNERS, IN **FOLLOWING WEEK(S)**

| | Learner Group | Step completion rates in the Following Week | | | |
|---|---|---|---|---|---|
| | | N | Mean Rank | M-W U | Sig. |
| Week 1 | Social | 4,918 | 35,802.17 | 32,081,722.5 | <.001 |
| | Non-social | 39,866 | 20,738.24 | | |
| | Social | 2,680 | 39,365.79 | 10,930,934.0 | <.001 |
| | Non-social | 42,104 | 21,312.12 | | |
| Week 2 | Social | 2,380 | 39,894.44 | 8,806,135.5 | <.001 |
| | Non-social | 42,404 | 21,410.17 | | |
| | Social | 1,767 | 41,050.15 | 5,037,444.0 | <.001 |
| | Non-social | 43,017 | 21,626.10 | | |
| Week 3 | Social | 1,787 | 40,669.65 | 5,756,549.0 | <.001 |
| | Non-social | 42,997 | 21,632.88 | | |
| | Social | 4,918 | 35,802.17 | 32,081,722.5 | <.001 |
| | Non-social | 39,866 | 20,738.24 | | |
| Week 4 | Social | 2,680 | 39,365.79 | 10,930,934.0 | <.001 |
| | Non-social | 42,104 | 21,312.12 | | |
| | Social | 2,380 | 39,894.44 | 8,806,135.5 | <.001 |
| | Non-social | 42,404 | 21,410.17 | | |
| Week 5 | Social | 1,767 | 41,050.15 | 5,037,444.0 | <.001 |
| | Non-social | 43,017 | 21,626.10 | | |
| | Social | 1,787 | 40,669.65 | 5,756,549.0 | <.001 |
| | Non-social | 42,997 | 21,632.88 | | |

## VIII. Discussions and Conclusions

This paper presents an exploratory study on a FutureLearn course aiming to investigate learners' learning versus social engagement. Thus, our main contributions with this paper are:

- A fine-grained temporal approach for learners' progress;
- Conducting research on the less-explored FutureLearn platform.
- Showing that, surprisingly, although completion numbers slowly dwindle, completions for active students tend to swell mid-course.
- Showing that, although social activities are at a much lower rate than learning activities, the two are good predictors for each other, both in the current week, as well as for consecutive weeks.

Our fine-grained temporal approach showcases the data-intensive analysis on *learning activities* and *social activities*, using statistical modelling, first responding to the research question on how *learning activities* and *social activities* change along the course. To answer the second research question, on the relationship between *learning activities* and *social activities*, we estimate the appropriateness to use predictive models. Here, we show that *social activities*, i.e., comments, could be used as parameters to predict whether a learner will complete most of the *steps* in the current and following week(s); similarly, *learning activities*, i.e., visiting a *step* and marking a *step* as complete, could be used as parameters to predict whether a learner will interact with peers, such as by posting comments in the current and following week(s).

Our results indicate that "rates" of *learning activities* are much higher than "rates" of *social activities*, suggesting that, although FutureLearn is built upon social learning pedagogy, the social features may be less popular than the learning activities. This can also indicate that people who start the courses with the primary objective of learning about a certain subject will be directly motivated by the pedagogical activities, independently of the social activities. However, this may be taken with a grain of salt, due to the limited social features available in FutureLearn.

Overall, our fine-grained temporal analysis showed that *learning activities* and *social activities* are good potential predictors of each other, and thus of utmost importance for designers and instructors of MOOCs, as well as developers of MOOC platforms: when designing a MOOC, it is important for the course designer to consider embedding effective social tasks, such as setting up discussion forums; when facilitating a MOOC, it is important for the course instructor to encourage learners to interact with peers, as this may further learning as well; when developing a MOOC platform, it is important for developers to develop sufficient social features.

In terms of future work, we plan to investigate engagement patterns of learners who enrolled and followed multiple runs of the same course. When analysing the dataset, we have already found that there was a certain percentage of learners appearing in several runs of the course, and they might be engaged differently in comparison with those learners who participated in only one run of the course. The results may help improve our predictive model, by reducing the "noise," as well as shed light on why some learners would repeat a course and its effects on their learning engagement and learning outcomes.